\documentclass{PoS}

\title{Heavy quark meson spectroscopy at CDF}

\ShortTitle{Heavy quark meson spectroscopy at CDF}

\author{\speaker{Felix WICK}%
         \thanks{On behalf of the CDF Collaboration.}\\
        University of Karlsruhe\\
        E-mail: \email{wick@ekp.uni-karlsruhe.de}}

\abstract{With growing datasets collected by the CDF II experiment, studies of the spectroscopy of mesons containing heavy quarks become more exciting. The CDF experiment has good capabilities in both charm and bottom sector. This capability allowed also to contribute to the study of the zoo of states called $X,Y,Z$. In this area we present a recent update of the mass measurement of $X(3872)$. The result $m(X(3872)) = 3871.61 \pm 0.16 \pm 0.19\,\mathrm{MeV/c}^2$ is currently the most precise measurement in the world. In addition, we report evidence for a new narrow resonance, $Y(4140)$, the first to be seen in the $J/\psi \phi$ decay mode, using $2.7\,\mathrm{fb}^{-1}$ of exclusive $B^+ \to J/\psi \phi K^+$ decays.}

\FullConference{European Physical Society Europhysics Conference on High Energy Physics\\
		 July 16-22, 2009\\
		 Krakow, Poland}

\begin{document}

\section{Introduction}

In recent years, several states with charmonium-like decays were discovered which do not fit properly into the established charmonium picture. These states, called $X,Y,Z$, are candidates for exotic mesons beyond the conventional quark-antiquark-model ($q\bar{q}$). Possible interpretations are quark-gluon-hybrids ($q\bar{q}g$), four-quark states ($q\bar{q}q\bar{q}$), molecular states composed of two usual mesons, or glueballs. In the following, two CDF analyses in this context are considered, a measurement of the $X(3872)$ mass in $J/\psi \pi^+ \pi^-$ decays \cite{XPaper} and an evidence for a narrow near-threshold structure in the $J/\psi \phi$ mass spectrum, called $Y(4140)$, using $B^+ \to J/\psi \phi K^+$ decays \cite{YPaper}.

Although the existence of $X(3872)$ is well established, its nature is still unclear. The two most popular models are a molecular state composed of $D^0$ and $\bar{D}^{*0}$ mesons and a four-quark state ($c\bar{c}q\bar{q}$). The molecular model is motivated by the closeness of the measured mass to the sum of the $D^0$ and $D^{*0}$ masses and could be rejected if $m(X(3872))$ was found larger than this sum. The four-quark state hypothesis predicts the existence of two distinct particles with different light-quark content and slightly different masses. Because of these predictions, an experimental examination of the $X(3872)$ mass and its structure allows to test the two hypotheses.

Consisting of two vector mesons, the final state $J/\psi \phi$ is a good channel to search for an exotic meson. Because the invariant mass in this channel is high enough for open charm decays, a charmonium state is very unlikely to be observed. Using exclusive $B^+$ decays to $J/\psi \phi K^+$, a strong background reduction can be achieved by exploiting the long $B$ meson lifetime and the identification of the additional kaon.

The data used for the reported analyses were collected by the CDF II experiment at the Tevatron proton-antiproton collider at Fermilab. CDF II is a multipurpose detector with an excellent tracking and mass resolution. Therefore it is well suited to perform the described measurements.

\section{$X(3872)$ Mass}

$X(3872)$ candidates are reconstructed in the decay channel $J/\psi \pi^+ \pi^-$ with $J/\psi \to \mu^+ \mu^-$. $\psi(2S)$ decays to the same final state and is used as a control sample. In order to separate signal from background, a neural network is trained by means of simulated $X(3872)$ events and the sidebands in the data mass spectrum. The most important input quantities for the neural network are the $Q$ value of the decay, the transverse pion momenta, the kinematic fit quality of the $X(3872)$ candidate and muon identification quantities. Subsequently, cuts on the output of this neural network and the number of candidates per event are performed which select around 34500 $\psi(2S)$ and 6000 $X(3872)$ signal events.

\begin{figure}
\centering
a)
\includegraphics[width=.34\textwidth]{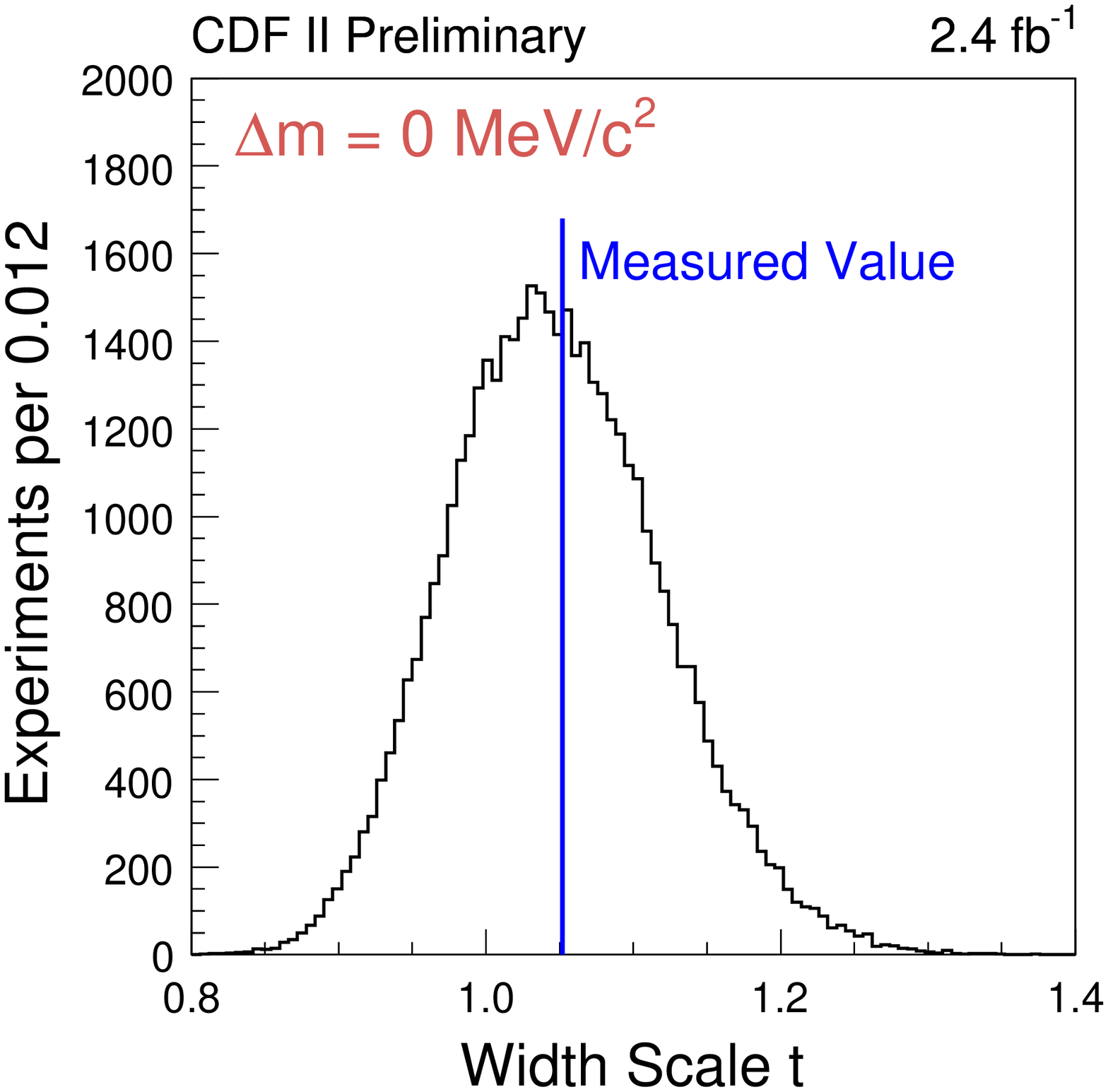}
\hspace{20mm}
b)
\includegraphics[width=.34\textwidth]{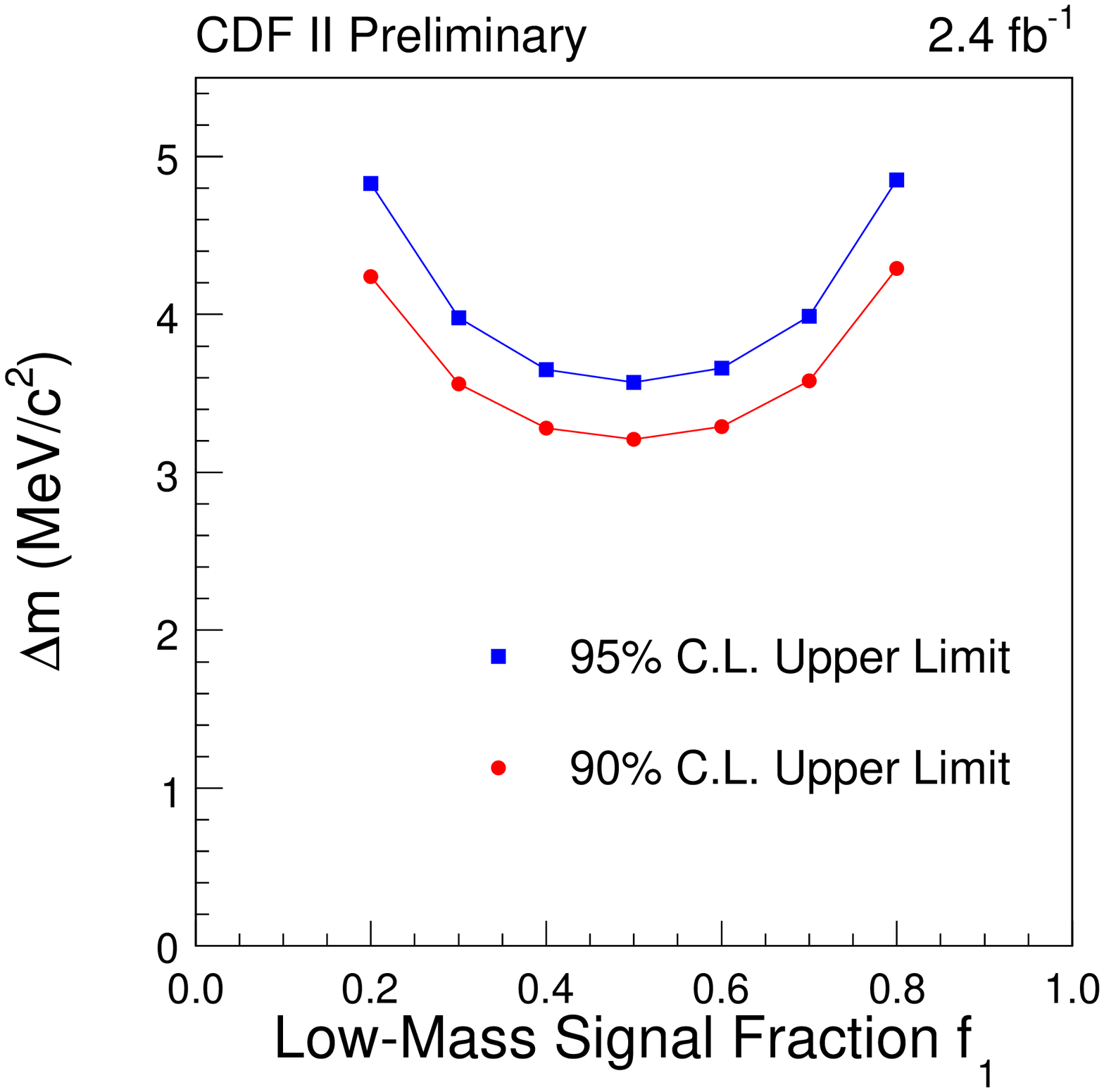}
\caption{(a) Distribution of the width scale $t$ from pseudoexperiments assuming a single $X(3872)$ state. Also shown is the value measured in data. (b) Upper limits on the mass difference between two possible states in dependence of the low-mass signal fraction.}
\label{Limit}
\end{figure}

In order to test whether there is evidence for more than one state, a binned maximum likelihood fit to the mass spectrum is performed, where the signal is described by a nonrelativistic Breit-Wigner function convolved with a Gaussian resolution and the combinatorial background is modeled by a second order polynomial. Thereby the resolution is obtained from MC simulations and the Breit-Wigner width is fixed to $1.34\,\mathrm{MeV/c^2}$ (average of \cite{XBelle} and \cite{XBabar}). But additionally, the overall width of the signal shape is scaled by a factor $t$ which serves as a free parameter in the fit. The parameter $t$ is also determined for ensembles of simulated experiments consisting of two $X(3872)$ states with given mass differences which are generated using the same fit model as in data but varying several quantities, like the Breit-Wigner width, according to their uncertainties. Figure \ref{Limit} (a) shows the comparison of the measured value $t$ with its distribution resulting from an ensemble of pseudoexperiments assuming a single state ($\Delta m = 0\,\mathrm{MeV/c^2}$). Here no evidence for two states can be found and an upper limit on their mass difference can be set by finding the value for which the scaling parameter $t$ of $90\%$ ($95\%$) of the pseudoexperiments is higher than the measured $t$. In figure \ref{Limit} (b) this limit is shown in dependence of the low-mass signal fraction $f_1$. For $f_1 = 0.5$ the limits are $m < 3.2\,\mathrm{MeV/c^2}$ and $m < 3.6\,\mathrm{MeV/c^2}$ at $90\%$ and $95\%$ confidence levels, respectively. This can be compared to the prediction of $\Delta m = (8 \pm 3)\,\mathrm{MeV/c^2}$ by the four-quark state model of Maiani et al.\cite{Maiani}.

\begin{figure}
\centering
a)
\includegraphics[width=.34\textwidth]{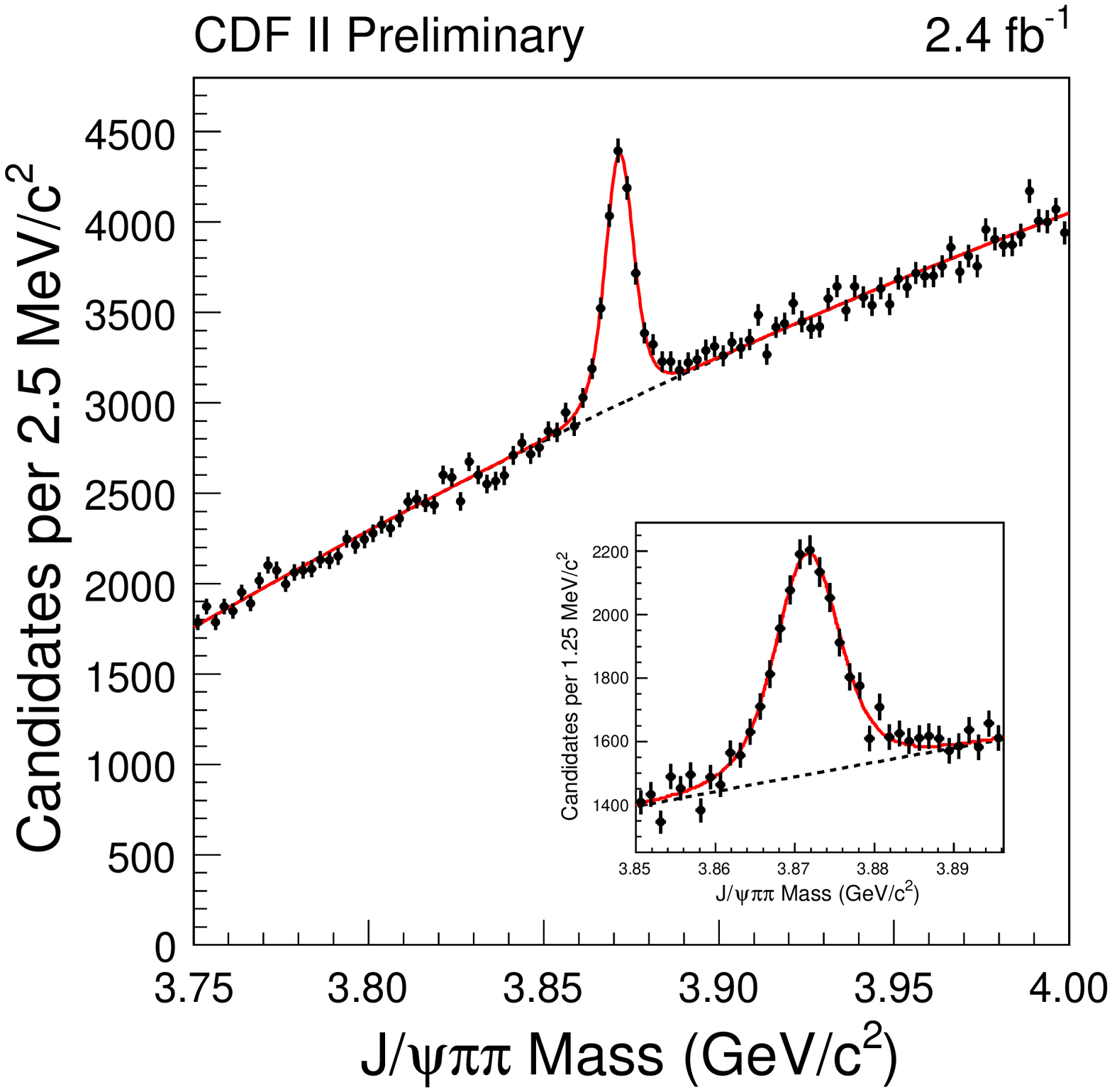}
\hspace{15mm}
b)
\includegraphics[width=.46\textwidth]{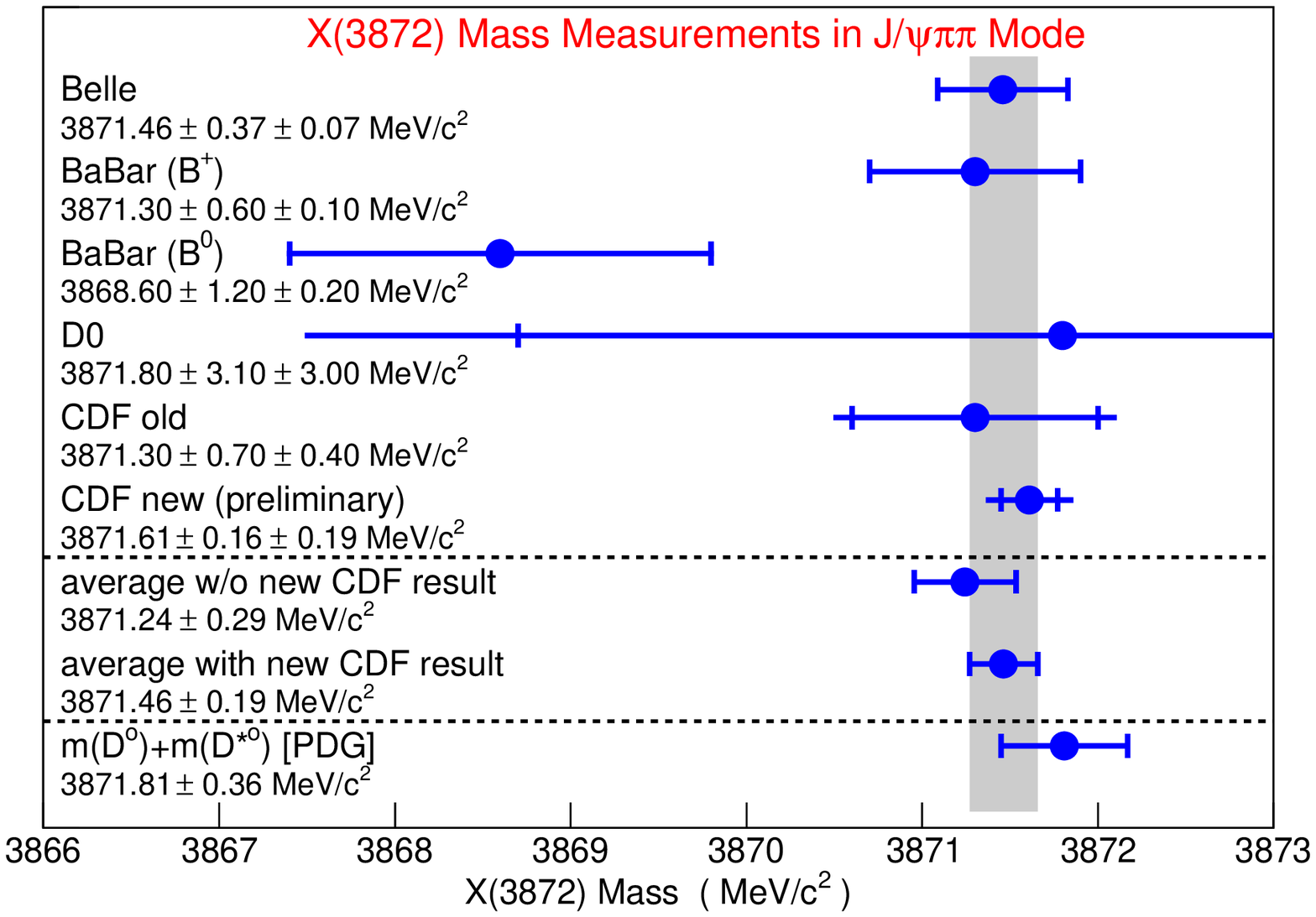}
\caption{(a) Invariant mass distribution of the $X(3872)$ candidates with the fitted function (solid red line) and the background model (dashed line). (b) $X(3872)$ mass measurements in the decay to $J/\psi \pi^+ \pi^-$.}
\label{XMass}
\end{figure}

Since the data are fully consistent with a single state, the mass of the $X(3872)$ can be measured by performing an unbinned maximum likelihood fit using the same fit model as described above. The result of this fit is shown in figure \ref{XMass} (a). The momentum scale uncertainty as main source of systematic errors can be estimated by means of the difference between the measured $\psi(2S)$ mass and the corresponding world average value. As can be seen in figure \ref{XMass} (b), the resulting value $m(X(3872)) = 3871.61 \pm 0.16 \pm 0.19\,\mathrm{MeV/c}^2$ is the most precise single measurement to date and lies below $m(D^0) + m(D^{*0})$. So the molecular model is still possible.

\section{Evidence for $Y(4140)$}

\begin{figure}
\centering
a)
\includegraphics[width=.34\textwidth]{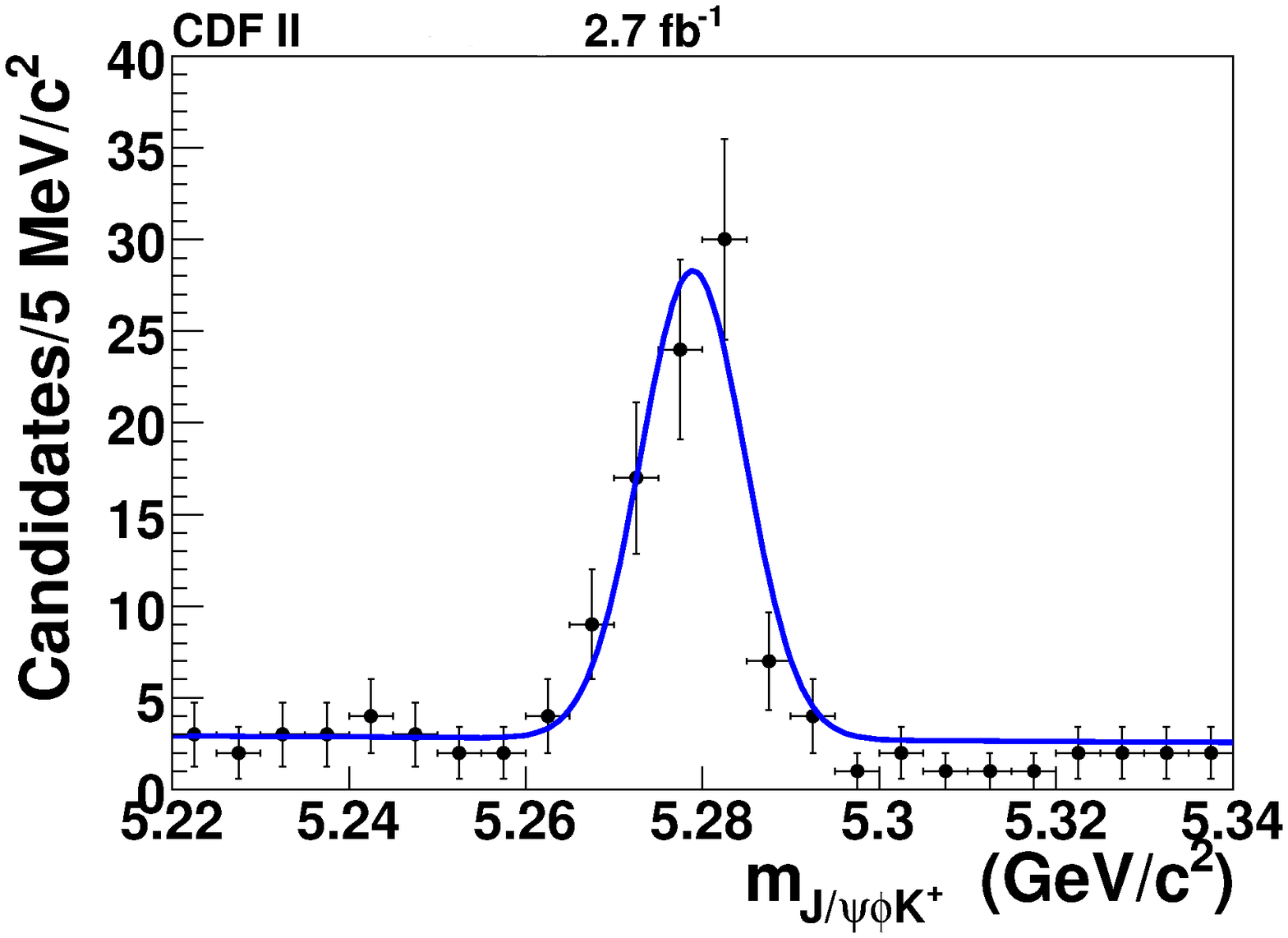}
\hspace{20mm}
b)
\includegraphics[width=.34\textwidth]{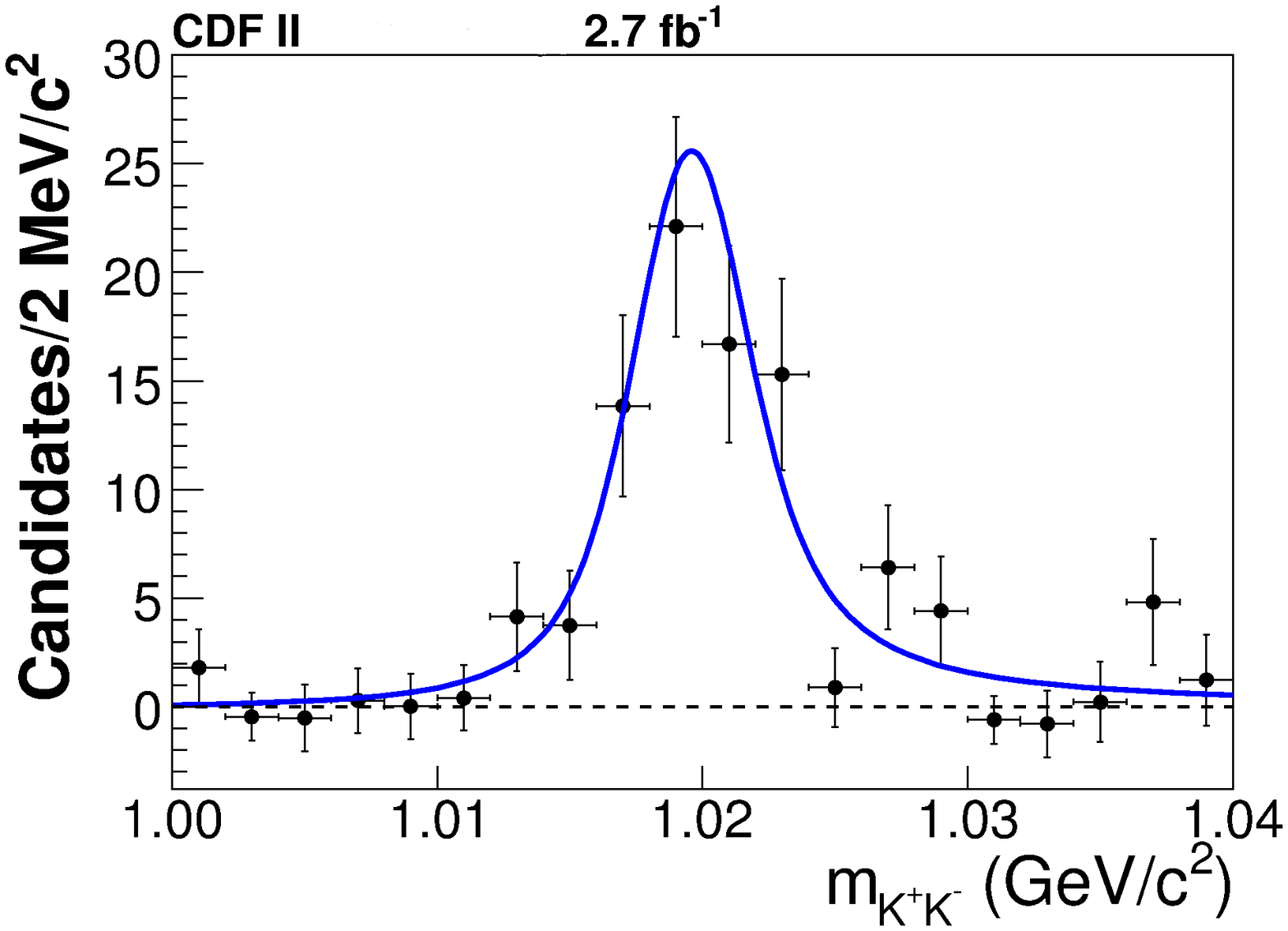}
\caption{Invariant mass distributions of (a) $J/\psi \phi K^+$ and (b) $K^+ K^-$ with the fitted functions (solid blue lines) as described in the text.}
\label{BSpectrum}
\end{figure}

In order to build $B^+ \to J/\psi \phi K^+$ candidates, first $J/\psi \to \mu^+ \mu^-$ and $\phi \to K^+ K^-$ candidates are reconstructed which are then combined with an additional charged track with kaon mass hypothesis. Thereby the reconstructed $J/\psi$ and $\phi$ masses are required to lie within $50\,\mathrm{MeV/c^2}$ ($J/\psi$) and $7\,\mathrm{MeV/c^2}$ ($\phi$) of the corresponding world average values. The combinatorial background can be reduced significantly by cutting on the $B^+$ decay length in the transverse plane ($L_{xy}(B^+)$) to exploit the long $B$ meson lifetime. In addition, a kaon identification quantity can be used to obtain a further background reduction. For that purpose, the information about the energy loss $dE$/$dx$ in the drift chamber and the information from the Time-Of-Flight detector are summarized in a log-likelihood ratio $LLR_\mathrm{Kaon}$. The cuts $L_{xy}(B^+)>500\,\mu\mathrm{m}$ and $LLR_\mathrm{Kaon}>0.2$ are chosen by optimizing the quantity $S/\sqrt{S+B}$, where $S$ and $B$ are the numbers of $B^+$ signal and background events, respectively. Figure \ref{BSpectrum} (a) shows the resulting $J/\psi \phi K^+$ invariant mass spectrum. A fit with a Gaussian signal and a linear background function yields $75\pm10$ signal events which is the largest sample to date. For the examination of the $J/\psi \phi$ spectrum, only candidates within $\pm 3 \sigma (17.7\,\mathrm{MeV/c^2})$ around the nominal $B^+$ mass are selected. Figure \ref{BSpectrum} (b) shows the $B^+$ sideband-subtracted $\phi \to K^+ K^-$ invariant mass spectrum without $\phi$ mass window requirement. The fit function is a $P$-wave relativistic Breit-Wigner convolved with a Gaussian to account for the detector resolution. As there is no significant background contribution, the $B^+ \to J/\psi K^+ K^- K^+$ final state is well described as $J/\psi \phi K^+$.

\begin{figure}
\centering
a)
\includegraphics[width=.34\textwidth]{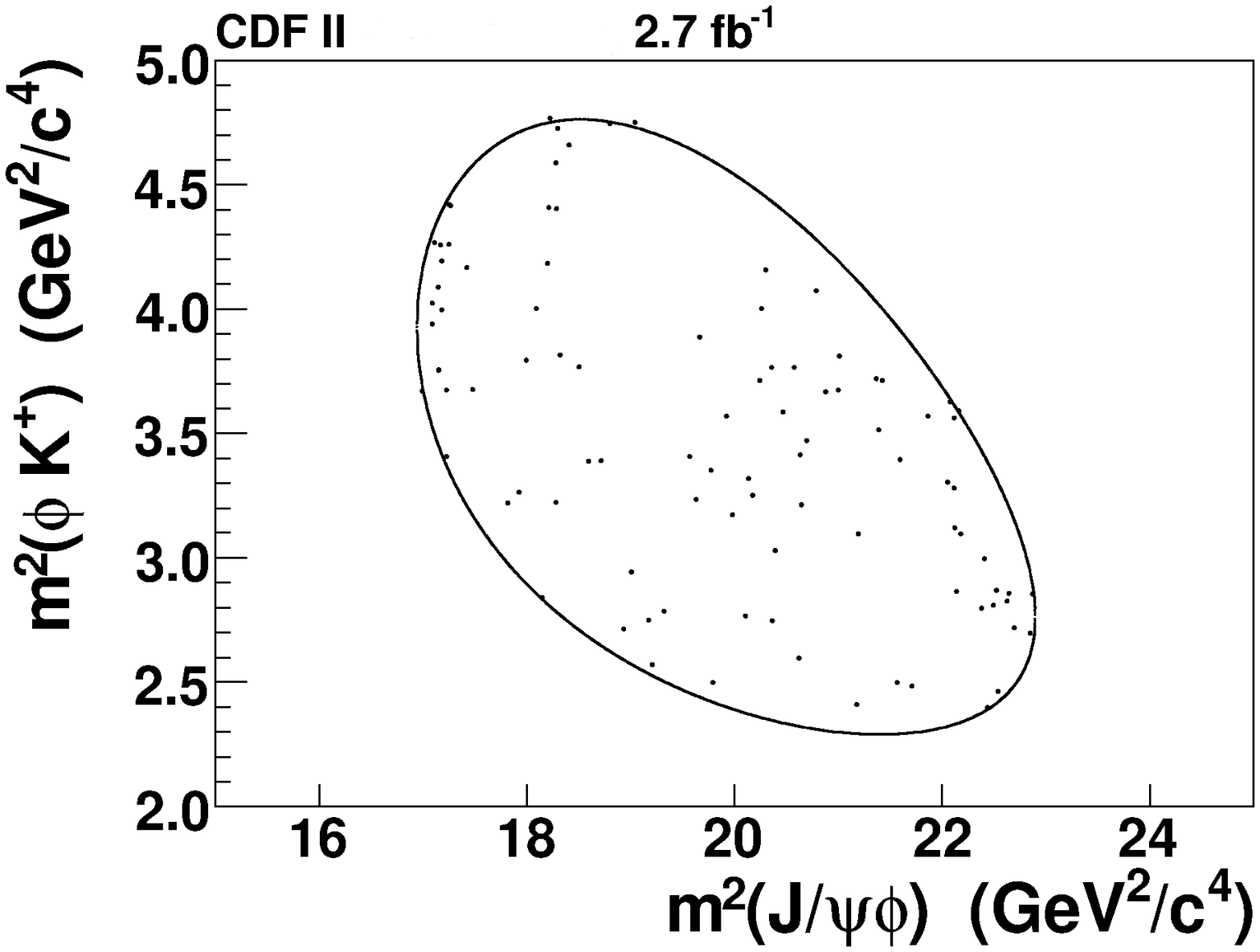}
\hspace{20mm}
b)
\includegraphics[width=.34\textwidth]{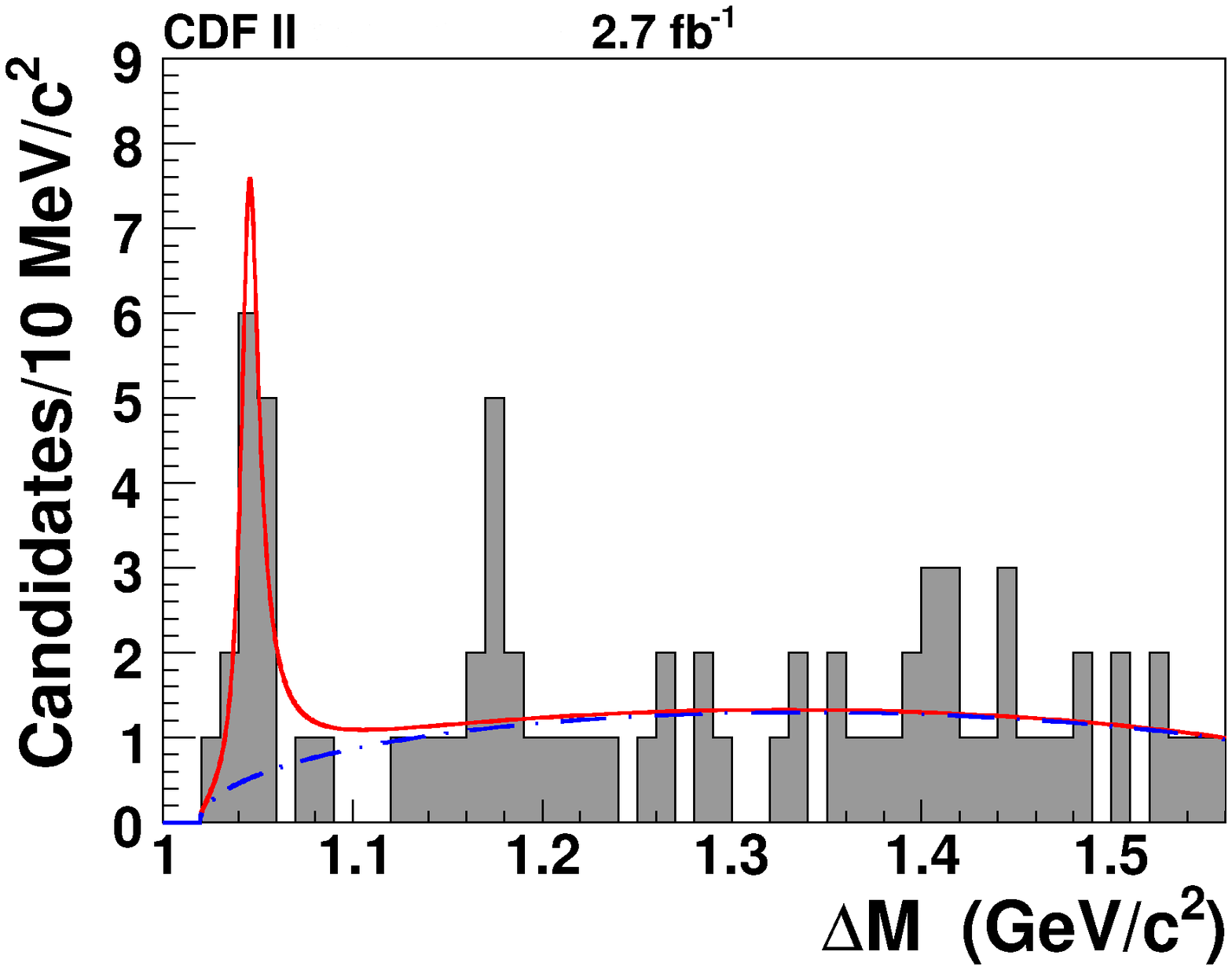}
\caption{(a) Dalitz plot of the final state $J/\psi \phi K^+$ in the $B^+$ mass window. The boundary shows the kinematically allowed region. (b) Distribution of the invariant mass difference $\Delta M = m(\mu^+ \mu^- K^+ K^-) - m(\mu^+ \mu^-)$ in the $B^+$ mass window with the fitted function (solid red line) and the background model (dashed blue line).}
\label{YSpectrum}
\end{figure}

In a Dalitz Plot of the considered decay (see figure \ref{YSpectrum} (a)), an enhancement near the threshold of $m^2(J/\psi \phi)$ can be seen. Figure \ref{YSpectrum} (b) shows the spectrum of the mass difference $\Delta M = m(\mu^+ \mu^- K^+ K^-) - m(\mu^+ \mu^-)$ using 73 events with $\Delta M < 1.56\,\mathrm{GeV/c^2}$. An unbinned maximum likelihood fit is performed, where the enhancement is described by the convolution of an $S$-wave relativistic Breit-Wigner function with a Gaussian resolution with the rms fixed to $1.7\,\mathrm{MeV/c^2}$ obtained from MC simulations, and the background is modeled by three-body phase space. The fit yields $14 \pm 5$ signal events and returns a mass difference $\Delta M = (1046.3 \pm 2.9)\,\mathrm{MeV/c^2}$ and a width of $(11.7^{+8.3}_{-5.0})\,\mathrm{MeV/c^2}$.

In order to estimate the probability of a creation of such a signal due to background fluctuations, simulations of the background distribution are performed and the number of trials which produce a signal with a log-likelihood ratio $-2 \ln{(\mathcal{L}_0/\mathcal{L}_{max})}$ of the null hypothesis fit and the signal hypothesis fit larger than the measured value are counted. Thereby the mass can be anywhere in the considered $\Delta M$ window and the width has to be larger than the detector resolution and smaller than ten times the observed width. In addition, the log-likelihood ratio in data is decreased by modeling the combinatorial background in the $B^+$ mass window separately as flat spectrum in $\Delta M$ instead of using pure three-body phase space background. This procedure leads to a significance of the enhancement of about $3.8 \sigma$.

With the world-average value for the $J/\psi$ mass, the mass and width of the observed structure are $(4143.0 \pm 2.9 \pm 1.2)\,\mathrm{MeV/c^2}$ and $(11.7^{+8.3}_{-5.0} \pm 3.7)\,\mathrm{MeV/c^2}$, respectively, where the systematic uncertainties are estimated by varying the fit model.

\section{Summary}

The presented measurement of the $X(3872)$ mass in its decay to $J/\psi \pi^+ \pi^-$ with a value of $(3871.61 \pm 0.16(stat) \pm 0.19(syst))\,\mathrm{MeV/c^2}$ is the most precise measurement to date. Additionally, an upper limit on the mass difference of two hypothetical $X(3872)$ states is derived.

In the $J/\psi \phi$ mass spectrum of exclusively reconstructed $B^+ \to J/\psi \phi K^+$ decays, evidence for a narrow near-threshold structure with a significance in excess of $3.8 \sigma$ is found. The measurements of the mass and width of this structure yield $(4143.0 \pm 2.9(stat) \pm 1.2(syst))\,\mathrm{MeV/c^2}$ and $(11.7^{+8.3}_{-5.0}(stat) \pm 3.7(syst))\,\mathrm{MeV/c^2}$, respectively.

\end{document}